\documentclass{imsart}

\RequirePackage{amsthm,amsmath,amsfonts,amssymb}
\RequirePackage[authoryear]{natbib}%% uncomment this for author-year citations
\RequirePackage[colorlinks,citecolor=blue,urlcolor=blue,backref=page,backref=page]{hyperref}
\RequirePackage{graphicx}

\pubyear{2024}
\volume{TBA}
\issue{TBA}
\firstpage{1}
\lastpage{1}

\startlocaldefs
\theoremstyle{plain}

\theoremstyle{definition}

\theoremstyle{remark}

\endlocaldefs

\begin{document}

\begin{frontmatter}
\title{Modeling EEG Spectral Features through Warped Functional Mixed Membership Models}
%\title{A sample article title with some additional note\thanksref{t1}}
\runtitle{Curve Registration for Functional Mixed Membership Models}
%\thankstext{T1}{A sample additional note to the title.}

\begin{aug}
%%%%%%%%%%%%%%%%%%%%%%%%%%%%%%%%%%%%%%%%%%%%%%%
%% Only one address is permitted per author. %%
%% Only division, organization and e-mail is %%
%% included in the address.                  %%
%% Additional information can be included in %%
\author[A]{\fnms{Emma}~\snm{Landry}\ead[label=e1]{emmalandry@ucla.edu}},
% \author[A]{\fnms{Second}~\snm{Author}\ead[label=e2]{second@somewhere.com}\orcid{0000-0000-0000-0000}}
\author[A] {\fnms{Damla}~\snm{\c{S}ent\"{u}rk}\ead[label=e2]{dsenturk@ucla.edu}}
\author[B] {\fnms{Shafali}~\snm{Jeste}\ead[label=e3]{}}
\author[C] {\fnms{Charlotte}~\snm{DiStefano}\ead[label=e4]{}}
\author[D] {\fnms{Abigail}~\snm{Dickinson}\ead[label=e5]{}}
\and
\author[A,E] {\fnms{Donatello}~\snm{Telesca}\ead[label=e6]{donatello.telesca@gmail.com}}
%%%%%%%%%%%%%%%%%%%%%%%%%%%%%%%%%%%%%%%%%%%%%%
%% Addresses                                %%
%%%%%%%%%%%%%%%%%%%%%%%%%%%%%%%%%%%%%%%%%%%%%%
\address[A]{Department of Biostatistics,
University of California, Los Angeles, CA 90095\printead[presep={\ }]{}}

\address[B]{Department of Neurology, Children's Hospital Los Angeles, Los Angeles, CA 90027 \printead[presep={\ }]{}}

\address[C]{Division of Clinical Psychology, Children's Hospital Los Angeles, Los Angeles, CA 90027\printead[presep={\ }]{}}

\address[D]{Center for Autism Research and Treatment, Semel Institute for Neuroscience, University of California, Los Angeles, CA 90095\printead[presep={\ }]{}}

\address[E]{Corresponding author\printead[presep={,\ }]{e6}}
\runauthor{E. Landry et al.}
\end{aug}

\begin{abstract}
A common concern in the field of functional data analysis is the challenge of temporal misalignment, which is typically addressed using curve registration methods. Currently, most of these methods assume the data is governed by a single common shape or a finite mixture of population level shapes. We introduce more flexibility using mixed membership models. Individual observations are assumed to partially belong to different clusters, allowing variation across multiple functional features. We propose a Bayesian hierarchical model to estimate the underlying shapes, as well as the individual time-transformation functions and levels of membership. Motivating this work is data from EEG signals in children with autism spectrum disorder (ASD). Our method agrees with the neuroimaging literature, recovering the 1/f pink noise feature distinctly from the peak in the alpha band. Furthermore, the introduction of a regression component in the estimation of time-transformation functions quantifies the effect of age and clinical designation on the location of the peak alpha frequency (PAF). 
\end{abstract}

% \begin{keyword}[class=MSC]
% \kwd[Primary ]{00X00}
% \kwd{00X00}
% \kwd[; secondary ]{00X00}
% \end{keyword}


\end{frontmatter}

\section{Introduction}
Functional data analysis is a branch of statistics concerned with the analysis of data that comes in the form of random curves. A key characteristic of such data is that it tends to display variation along both the $x$ and $y$ axes, respectively referred to as phase and amplitude variation. This manuscript introduces a Bayesian curve registration method to account for phase variation in normalized data arising from mixed membership models-- where individuals partially belong to multiple clusters associated with distinct functional features.

Our method is motivated by a case study of resting-state electroencephalography (EEG) measured in a cohort of children with Autism Spectrum Disorder (ASD) and age-matched Typically Developing (TD) children (Fig.~\ref{fig:data}). This data was first introduced, described and analyzed by \citet{Dickinson2018}. In particular, the focus is on spectral analysis in the alpha band (6-12 Hz), within which a noticeable peak occurs. The Peak Alpha Frequency (PAF), defined to be the frequency at which such a peak occurs for an individual, as well as the power associated with the peak,  can serve as neuromarkers for ASD, as well as multiple other neurodevelopmental disorders.

As a result, the study of the alpha band for resting state EEG has been of notable interest in the field of neuroscience. \citet{Deiber2020} conduct a study on adults with Attention-Deficit  Hyperactivity Disorder (ADHD), in which the control group is seen to display higher resting-state alpha power, that is, more prominent peaks. In the case of patients suffering from schizophrenia, \citet{Ramsay2021} note a slower PAF than the controls, i.e. the peak occurs at slower frequencies. Focusing on ASD, \citet{Edgar2015} find a difference in the relationship between age and PAF. Analysis of the data for our case study of interest agrees with those results, noting that TD children display an increase in PAF with age, which their ASD counterparts don't \citep{Dickinson2018}. Other authors in the statistical literature have previously tackled the analysis of this data \citep{Scheffler2019, Shamshoian2022,Marco2024}, but their analyses have not formally considered issues of phase variability to account for mismatch in timing of features across individuals.

%With such considerations, we now provide an overview of approaches developed for tackling the aforementioned problem of phase variation in FDA. A common example motivating the need for alignment, or registration, methods is the study of growth curves in children, where differences in age of development across individuals lead to misaligned data. Accounting for those differences in timing allows to compare other characteristics more effectively. 
The registration problem has been extensively explored in the study of functional data. In the frequentist literature, one of the earliest approaches is landmark registration \citep{Gasser1995}, where certain features of the data are identified to match their timings. \citet{Wang1997, Wang1999} consider alignment through dynamic time warping, where the shift between two functions is optimized through a cost function. \citet{Ramsay1998} introduce the Procrustes method for alignment, using an iteratively updated cross-sectional mean as a target to estimate time-transformation functions. Many authors have also considered more flexible methods that use self-modelling warping functions, such as through shape-invariant models (SIM) \citep{ronn2001}, or through B-spline basis estimation \citep{Gervini2004}. More recently, \citet{Srivastava2011} take a geometric approach borrowing from shape analysis methodology and use the Fisher-Rao metric as a distance measure for registration. 

The work of \citet{Telesca2008} introduces the first method that addresses the curve registration problem in the Bayesian framework. They present a hierarchical model to formally account for amplitude and phase variation. For some other recent development in Bayesian curve registration, see \citet{Claeskens2010}, \citet{Cheng2016}, \citet{Bharath2020}, \citet{Matuk2022}.

While our contribution is novel in that we consider the framework of mixed membership models, multiple authors have allowed for curve registration methods to incorporate the concept of groups, or features. In particular, several strategies for clustering of functional data while addressing the alignment problem have been developed \citep{Liu2009, Zhang2014, Tang2023}. Other approaches consider the case where data may vary in multiple functional directions, such as \citet{Kneip2008} who address it through functional principal components, or \citet{Earls2017} that combine curve registration with factor analysis.

In this paper, we extend Bayesian hierarchical curve registration introduced by \citet{Telesca2008} to mixed membership models \citep{Erosheva2004}.This class of models assumes the existence of a finite number of subpopulations, to which each individual belongs to varying degrees. \citet{Marco2024} introduce mixed membership to the field of functional data analysis, where the authors exploit the multivariate Karhunen-Loève theorem to extend finite Gaussian process (GP) mixture models. Our proposed model addresses the issue of phase variability, which arises from the fact that individual functions may operate on a distorted timescale, leading to misalignment of functional data in physical time \citep{Ramsay1997}. We leverage the work of \citet{Brumback2004} to model the time-transformation functions in an unconstrained fashion and transform them so that shape restrictions are satisfied. This allows for flexible modeling, including covariate adjustments, and the estimation through B-splines is conveniently incorporated into MCMC sampling. 

The remainder of the paper is organized as follows. Section~\ref{methods} introduces the curve registration model for the mixed membership framework-- including discussion of prior choices, identifiablity concerns, and posterior inference details. Section~\ref{applications} illustrates our method in practice. In section~\ref{simulationstudy} we present a simulation study to assess model fit, and explore the effects of semi-supervision through partial labelling of observations. Section~\ref{casestudysec} applies the model to electroencephalography (EEG) data. Finally, section~\ref{discussion} provides a critical discussion of our methodology.

\section{Methods} \label{methods}
\subsection{Joint registration and mixed membership modeling of peak alpha frequency curves} \label{modeldescription}

In general, mixed membership models may include any number of functional features or pure membership functions. Following \cite{Marco2024} and motivated by our neuroimaging case study,  this paper focuses on curve registration for a mixed membership model with two functional features. Restriction to this case allows for scientifically meaningful inference, while operating in a statistically-sound and easily interpretable setting. We make clear how mixed membership is formalized for misaligned functional data in the following. 

Let $Y_i(t)$ denote the unregistered function corresponding to the EEG spectral density for subject $i$, $(i =1,2,...,N)$ and $t  \in \mathcal{T} = [t_1,t_n]$. For each curve, we introduce a time transformation function $h_i(t): \mathcal{T} \rightarrow \mathcal{T}$, defining a a subject-specific ``stochastic schedule'', and a shape function $\mu_i(t):\mathcal{T} \rightarrow \mathbb{R}$, indexing the expected behavior of each function for a fixed evaluation schedule $t\in \mathcal{T}$, so that  we may assume $E\left[Y_i(t)\right] = \mu_i\left[h_i(t)\right]$. Although perfectly reasonable as a data generation scheme, likelihood identifiability of both $h_i()$ and $\mu_i()$ requires that specific anchoring assumptions are made about these functions. Most commonly, for example, one requires full shape homogeneity, s.t. $\mu_i(t) = \mu(t),\quad \forall 
 i \in \{1,\ldots, N\}$, discounting simple scale and level heterogeneity \citep{Telesca2008, Brumback2004}. Although seemingly innocuous, the notion of shape homogeneity is often too restrictive in applications, as real data are seldom faithful to a simple shape-invariant transformation. The first innovation introduced in this manuscript is to allow the shape function $\mu_i()$ to exhibit subject-specific variation as it partially expresses two pure membership processes, which we denote with $f_1(\cdot)$ and $f_2(\cdot)$.   

Let $c_i$ denote a curve-specific random intercept, and $\pi_i\in [0,1]$ define a mixing proportion, quantifying the level of membership to the first pure membership process. Individual shapes are then assumed to follow the mixed membership process below:  
\begin{equation}
    \mu_i(t)= c_i+  \pi_{i} f_1(t) + (1-   \pi_{i})f_2(t), \quad \forall i \in \{1, ..., N \}.
\end{equation}
%In the above, $c_i$ denotes the individual linear scale parameter, which provides flexibility for differing intercepts. $\boldsymbol \pi_i = [\pi_{i}, 1-   \pi_{i}]$ is the mixed membership vector, which denotes the individual level of membership to each feature, where 
%$ \pi_{i} \in [0,1], \quad   \forall i \in \{1, ..., N \}$.
Remembering $h_i(t)$ denotes the time-transformation function for subject $i$, the assumed sampling model for our data follows the familiar structure below: 
\begin{equation} \label{mainmodel}
    Y_i(t)= \mu_i[h_i (t)] + \epsilon_i(t), \quad  \forall i \in \{1, ..., N \};
\end{equation}
where the error terms $\epsilon_i(t)$ are assumed to be independent and normally distributed, such that $\epsilon_i(t) \stackrel{\text{iid}}{\sim} \mathcal{N}(0,\sigma^2_\epsilon)$ $ \forall i \in \{1, ..., N \}, \quad t \in \mathcal {T}. $

In the following we represent all functional components using B-spline basis systems \citep{Boor2001}. Other representations are possible, e.g. using Gaussian Processes, but are not discussed in this article. 

%$f_k$'s (where $k =1, 2$), denote the shape functions corresponding to each cluster, or subpopulation. 

We model the pure membership processes $f_1(\cdot)$  and $f_2(\cdot)$ using B-spline basis functions. Letting $(\kappa_1, ..., \kappa_p)$ denote the set of interior knots partitioning $\mathcal{T}$, and using a piecewise cubic polynomial, we denote with $ \boldsymbol {\mathcal B}_s(t)$  the $K = p +4$ - dimensional B-spline design vector evaluated at time $t$. Each pure membership feature can then be represented as the linear combination $f_k(t)=  \boldsymbol{\mathcal B}_s(t)^\prime \boldsymbol \gamma_k$, where $\boldsymbol \gamma_k$ is the vector of spline coefficients for  feature $k$.

The individual time-transformation functions $h_i(t)$ are similarly defined using B-splines. We once again use piecewise-cubic polynomials and consider $h$ interior knots $(\omega_1, ..., \omega_h)$. Letting $ \boldsymbol {\mathcal B}_w(t)$ denote the $Q = h + 4$ - dimensional vector of B-spline basis at time $t$, the time-transformation functions are given by $h_i(t)=\boldsymbol {\mathcal B}_w(t)^\prime \boldsymbol \phi_i$. $\boldsymbol \phi_i \in \mathbb R^Q$ is the vector of time-transformation spline coefficients for subject $i$. These curve-specific functions account for differences in timing that result in shifted location of features. It is hence natural to impose monotonicity, ensuring no time-reversing; and the image constraints $h_i(t_1) = t_1$, $h_i(t_n) = t_n$ such that the transformed times remain in the original time domain $\mathcal T$. We provide more details on how these constraints are imposed in practice in section \ref{warpingsubsection}.

Because warping acts on multiple shapes, we allow for the effect of warping to differ across features. Rather than estimating freely two time transformation functions per subject, we %estimate $h_i$ as described above using B-splines and let
assume homogeneity of the warping shape across features, but allow for varying degrees of shape deformation as follows:
\begin{equation} \label{ttrho}
    h_{i,1} (t) = \rho_1( h_i(t) - t) + t, \quad h_{i,2} (t) = \rho_2( h_i(t) - t) + t, \quad \rho_1, \rho_2 \geq 0.
\end{equation}
The parameters $\rho_k$ shrinks the time-transformation functions towards the identity for values less than 1, and intensifies the effect of warping when larger than 1. To ensure identifiability, and without loss of generality, we set $\rho_1 = 1$, reducing equation (\ref{ttrho}) to 
\begin{equation}
    h_{i,1} = h_i(t), \quad h_{i,2} =  \rho( h_i(t) - t) + t, \quad \rho \geq 0.
\end{equation}

Combining all of the above, we can then write 
\begin{equation*}
    m_i(t) \equiv \mu_i(h_i(t)) = c_i + \pi_i \boldsymbol {\mathcal B}_s(\boldsymbol {\mathcal B}_w(t)^\prime \boldsymbol \phi_i)^\prime \boldsymbol \gamma_1 + (1 - \pi_i)\boldsymbol {\mathcal B}_s(\rho[\boldsymbol {\mathcal B}_w(t)^\prime \boldsymbol \phi_i-t] + t)^\prime \boldsymbol \gamma_2
\end{equation*}
and the likelihood for the model is given by
\begin{equation} \label{continuslikelihoood}
    Y_i(t) \mid c_i, \pi_i, \boldsymbol \gamma_1, \boldsymbol \gamma_2, \boldsymbol \phi_i, \rho,\sigma^2_\epsilon \sim \mathcal{N}\big(m_i(t), \sigma^2_\epsilon\big),\quad  \forall i \in \{1, ..., N \}.
\end{equation}
In practice, functional observations are recorded discretely on a finite evaluation grid. We assume that measurements for subject $i$ are obtained at times $t_{i,1}, ..., t_{i, n_i}$, and can hence represent the data as the vectors $\boldsymbol Y_i \in \mathbb R^{n_i}, i = 1,..., N$. We can rewrite (\ref{continuslikelihoood}) using the multivariate normal distribution as
\begin{equation}
    \boldsymbol Y_i \mid c_i, \pi_i, \boldsymbol \gamma_1, \boldsymbol \gamma_2, \boldsymbol \phi_i, \rho,\sigma^2_\epsilon \sim \mathcal{N}_{n_i}\big(\mathbf m_i, \sigma^2_\epsilon \boldsymbol I _{n_i}\big), \quad \forall i \in \{1, ..., N \},
\end{equation}
where $\mathbf m_i$ is the vectorization of $m_i(t)$; that is, 
\begin{equation*}
    \mathbf m_i = c_i \boldsymbol 1_{n_i}+ \pi_{i} \boldsymbol {\mathcal B}_s(\boldsymbol {\mathcal B}_w(\boldsymbol t_i)^\prime \boldsymbol \phi_i)^\prime \boldsymbol \gamma_1 + (1- \pi_{i}) \boldsymbol {\mathcal B}_s(\rho[\boldsymbol {\mathcal B}_w(\boldsymbol t_i)^\prime \boldsymbol \phi_i-\boldsymbol t_i] + \boldsymbol t_i)^\prime \boldsymbol \gamma_2.
\end{equation*}
In the following Section we discuss covariate adjustment, by introducing a regression model for the explanation of phase variation. 

\subsection{Time-Transformation Function Regression} \label{warpingsubsection}
In this section, we discuss the estimation of the time-transformation functions in detail, and incorporate regression in order to leverage the effect of covariates on individual timings. As noted previously, the time-transformation functions need to satisfy several conditions in order for them to have sensical interpretation. We ask for them to be monotonically increasing, that is, $h_i(t_j) < h_i(t_k)$ for $t_j <t_k$, where $t_j, t_k \in \mathcal{T}$. \cite{Brezger2008} use the properties of B-spline derivatives to demonstrate that a sufficient condition for monotonicity is ordered spline coefficients. Hence, in our setting, we ask that 
\begin{equation}
    \phi_{i,1} < \ldots < \phi_{i,Q}, \quad  \forall i \in \{1, ..., N \}.
\end{equation}
Furthermore, it is desirable for the start and end points of the stochastic warped time window to correspond to the sampling time boundaries. The image constraint is imposed through the spline coefficients as well,
\begin{equation}
    \phi_{i,1} = t_{i,1}, \quad \phi_{i, Q} = t_{i, n_i}, \quad \forall i \in \{1, ..., N \}.
\end{equation}
Since the first and last entry are fixed, the estimation of time-transformation function only requires work with $ Q^* = Q - 2$ parameters for each subject. We let $\tilde {\boldsymbol\phi}_i = [\phi_{i,2}, ..., \phi_{i, Q-1}]^\prime $ denote the vector of free parameters for subject $i$, the object of interest for inference.

Several approaches have been explored in the Bayesian curve registration literature to tackle sampling with such constraints. 
These include sampling from a truncated support \citep{Telesca2008, Telesca2012}, exploiting the Riemannian geometry of warping functions and operating in a tangent space \citep{Lu2017}; or constructing a point process for increments \citep{Bharath2020, Matuk2022}. Because we are interested in explaining phase variability using covariate information, We follow the approach of \cite{Brumback2004} %, by operating in an unconstrained space for sampling, and applying 
and work with the Jupp transformation \citep{Jupp1978} to ensure monotonicity of the coefficients. Specifically, given a $Q$-dimensional vector $\boldsymbol{\phi}_i$, with ordered elements. s.t. $\phi_{i,1} < \phi_{i,2} < \cdots < \phi_{i,(Q-1)} < \phi_{i,Q}$, the Jupp transformation establishes a bijection, s.t. $\boldsymbol{\eta}_i = \mbox{Jupp}(\boldsymbol{\phi}_i) \in \mathbb{R}^Q$, unconstrained. Conversely, for any vector $\boldsymbol{\eta}_i \in \mathbb{R}^Q$, $\boldsymbol{\phi}_i = \mbox{Jupp}^{-1}(\boldsymbol{\tilde \eta}_i)$, with ordered elements. In our example, considering domain constraints, we operate with the unconstrained vector $\boldsymbol{\tilde \eta}_i\in \mathbb{R}^{Q^*}$, s.t. the warping coefficients are represented as $\boldsymbol{\phi}_i = \text{Jupp}^{-1}([0,\boldsymbol{\tilde \eta_i},1])$. Finally, denoting with $\boldsymbol X_i \in \mathbb{R}^{l}$, a vector  of $l$ covariate values for subject $i$, phase regression is encoded in the following multivariate linear model 
\begin{equation}
    E\left[\boldsymbol {\tilde \eta}_i\right] = \boldsymbol{ \tilde{\Upsilon}} + \mathbf B ^\prime \boldsymbol X_i, \quad \forall i \in \{1, ..., N\};
\end{equation}
where, $\boldsymbol {\tilde \Upsilon} \in \mathbb{R}^Q$ in the vector $\boldsymbol \Upsilon = [0, \boldsymbol {\tilde \Upsilon}, 1]^\prime $ correspondinds to the identity spline coefficients, s.t. $h_i(t) = \boldsymbol {\mathcal B}_w(t)^\prime \boldsymbol \Upsilon = t, \quad \forall t\in [0,1]$, and $\mathbf B \in \mathbb R^{ l \times Q^* }$ is a matrix of phase regression coefficients.

%we sample $Q^*$ free parameters $\tilde {\boldsymbol \eta}_i$ from the Gaussian distribution (more details in Section \ref{sectionpriors}), and the time-transformation coefficients are then obtained as $\boldsymbol \phi_i = \text{Jupp}^{-1}([0, \tilde {\boldsymbol\eta}_i, 1]^\prime)$. 
We provide the definition and details for computation of the Jupp and inverse Jupp transformations in the supplementary material. An advantage of the selected approach is that operation in the unconstrained Jupp space provides a convenient framework for regression on time-transformation functions. Indeed, the model can be easily extended such that the $\tilde {\boldsymbol \eta}_i$ are estimated with the inclusion of covariates, which can help give insights into the effect of different subject characteristics on individual temporal misalignment. %More details regarding the parametrization can be found in section \ref{sectionpriors}.

\subsection{Prior Distributions} \label{sectionpriors}
We now detail the specifics for the sampling model described in the previous sections, in particular the prior choices made for the parameters. Starting with the individual level intercepts, an intuitive and natural choice of prior is the Gaussian distribution with a hierarchical structure for variance estimation, such that
\begin{equation}
    c_i \mid \sigma^2_c \stackrel {\text{iid}}\sim \mathcal{N} (0, \sigma^2_c), \quad  \forall i \in \{1, ..., N \}.
\end{equation}
We ask for the intercepts to be centered at zero, so as to have them be interpreted as individual departure from the level determined by the shape features. As detailed earlier, these are modeled using B-splines. We consider the vector $\boldsymbol \gamma = [\boldsymbol \gamma_1, \boldsymbol \gamma_2]^\prime \in \mathbb{R}^{2K}$, which consists of the stacked vectors of spline coefficients for each feature. We assume a multivariate Gaussian prior
\begin{equation}
    \boldsymbol \gamma \mid \boldsymbol \Sigma_\gamma \sim \mathcal{N}_{2K}(\boldsymbol 0, \boldsymbol \Sigma_\gamma ).
\end{equation}
Assuming zero mean allows to not make any assumptions a priori regarding the shapes represented by each feature. The covariance structure is addressed with more care. Rather than treating $\boldsymbol \Sigma_\gamma$ as a fully unconstrained matrix whose entries need to be learned, we follow the penalized B-spline approach introduced by \citet{Lang2004}, who assume a first-order random walk shrinkage prior for the spline coefficients. Under this assumption, a penalization matrix for the covariance
\begin{equation}
\label{precisionpenalization}
    \boldsymbol \Omega = \begin{bmatrix}
    2 & - 1 & & & &0 \\
    -1 & 2 & -1 & \ddots & & & \\
    0 & -1 & \ddots & \ddots & \ddots & \\
    & \ddots & \ddots & \ddots & -1 & 0 \\
    & & \ddots & -1 & 2 & -1\\
    0 & & & 0 & -1 & 1
\end{bmatrix}
\end{equation}
is introduced. Assuming independence of the shape of the features, and introducing variance parameters $\lambda_1, \lambda_2$, the covariance is then given by the block diagonal matrix 
\begin{equation}
    \boldsymbol \Sigma_\gamma = \begin{bmatrix}
        \lambda_1 \boldsymbol \Omega^{-1} & 0 & \\
        0& \lambda_2 \boldsymbol \Omega ^{-1}
    \end{bmatrix}
\end{equation}
The two variance parameters need to be estimated, rather than the 4$K^2$ entries. This strategy, reduces sensitivity to the choice of $K$ by penalizing finite differences between adjacent coefficients.  

%Given the constraints the coefficients are subject to, earlier exposed in section \ref{warpingsubsection}, we operate fully in the unconstrained space for the estimation of the time-transformation function coefficients. Assuming an independent covariance structure, we then have that 

Considering the inherent smoothness associated with the ordering constraints in sec.~\ref{warpingsubsection}, individual time-transformation coefficients are assigned a regression prior in the Jupp space, s.t.  
\begin{equation}
    \boldsymbol {\tilde \eta}_i \mid \sigma^2_\eta \stackrel{\text{iid}}{\sim} \mathcal{N}_{Q^*}(\boldsymbol{ \tilde{\Upsilon}} + \mathbf B ^\prime \boldsymbol X_i , \sigma^2_\eta\boldsymbol I_{Q^*}), \quad \forall i \in \{1, ..., N\}.
\end{equation}
%\begin{equation}
%    \boldsymbol {\tilde \eta}_i \mid \sigma^2_\eta \stackrel{\text{iid}}{\sim} \mathcal{N}_{Q^*}(\boldsymbol {\tilde \Upsilon}, \sigma^2_\eta \boldsymbol I_{Q^*}), \quad \forall i \in \{1, ..., N\}.
%\end{equation}
%In the above, $\boldsymbol {\tilde \Upsilon}$ is such that $\boldsymbol \Upsilon = [0, \boldsymbol {\tilde \Upsilon}, 1]^\prime $ is the vector corresponding to the identity spline coefficients. In other words, it is the Jupp-transformed B-spline coefficient vector of dimension $Q$ corresponding to the identity time-transformation $h(t) = t$ (see the supplementary material for more detail). 
The ensuing prior for the $\boldsymbol \phi_i$', while not available as a standard distribution, remains centered around the parameter corresponding to no warping effect, and variation around it can be interpreted as stochastic time growing faster than physical time, or the opposite. Crucially, the covariate vector $\boldsymbol X_i \in \mathbb{R}^{l}$ does not include an intercept term as we require centering of the time-transformation functions around the identity. %This parametrization provides a convenient and natural framework for the incorporation of covariates. Indeed, the prior for the unconstrained parameters can be modified to become

The matrix of regression coefficients $\mathbf B \in \mathbb R^{ l \times Q^* }$ is %the matrix of regression coefficients that is 
assumed to follow the Matrix Normal Distribution
\begin{equation}
    \mathbf B \mid \sigma^2_\eta \sim \mathcal{MN}_{l,Q^*}(\mathbf{B}_0, \mathbf V_{B}, \sigma^2_\eta \mathbf I_{Q^*}).
\end{equation}
For both simulations and the case study, we let $\boldsymbol B_0 = \boldsymbol 0$, and take a g-prior for $\mathbf B$ by setting $\mathbf V_B = g(\boldsymbol X ^\prime \boldsymbol X)^{-1}$, where $\boldsymbol{X}\in \mathbb{R}^{N\times \ell}$ is the usual design matrix stacking individual covariate vectors.
%$$
%\mathbf X = \begin{bmatrix}
%    \mathbf{X}_1^\prime\\
%    \vdots \\
%    \mathbf{X}_N ^\prime
%\end{bmatrix} \in \mathbb R ^{N \times l}.$$
The warp scaling parameter $\rho$, modulating the effect of warping for the second feature, is assumed to follow a Gamma distribution,
\begin{equation}
    \rho \sim \text{Gamma}(a_\rho, b_\rho).
\end{equation}
%In practice, we choose hyperparameters that lead to mass highly concentrated near zero. While this choice is motivated by our case study, where we expect little warping effect to the $1 / f$ noise, in practice we find little sensitivity to the prior and convergence to the truth is achieved in %Since the mixed membership components lie on a simplex, a Dirichlet prior is a natural choice
%\begin{equation}
%  \boldsymbol \pi_i  \stackrel {\text{iid}}\sim \text{Dir}(\boldsymbol \alpha), \quad \forall i \in \{1,..., N\}. 
%\end{equation}
%A hierarchical structure could be introduced to let $\boldsymbol \alpha$ be a hyperparameter to be estimated, however we find that keeping a fixed value yields good performance both in our simulation study and in the analysis. 
Finally, mixed membership parameters $\pi_i$ are assumed iid and assigned a Beta distribution, s.t.
$$\pi_i \stackrel{\text{iid}}{\sim}  \text{Beta}(\alpha, \alpha).$$
All variance hyperparameters are learned from data. It is convenient to exploit conjugacy, and Inverse-Gamma distributions are hence assumed a priori for those, as well as for the random error variance,
\begin{equation}
    \sigma^2_c \sim \text{IG}(a_c, b_c), \quad \lambda_k \stackrel {\text{iid}}\sim \text{IG}(a_\lambda, b_\lambda), \quad \sigma^2_\eta \sim \text{IG}(a_\eta, b_\eta), \quad \sigma^2_\epsilon \sim \text{IG}(a_\epsilon, b_\epsilon).
    \end{equation}

\subsection{Constraints and Likelihood Identifiability} \label{identifiability}
Mixed membership models with joint alignment are not likelihood identifiable, unless constraints are introduced to eliminate sampling ambiguities. %Some that are common to curve registration while others arise from the mixed membership framework. 

Starting with identifiability issues typically associated with curve alignment, we consider the following constraints.   First, we impose the sum-to-zero constraint $\sum _{i=1}^N c_i =0 $ on the intercepts. This ensures that the level is fixed by the intercept and is not confounded with the shape features. A similar concern surrounds warping, which, without constraints, could be captured in the estimation of the features. We address this issue by centering the time-transformation functions around the identity. We operate in the unconstrained space, and set $\sum_{i=1}^N \boldsymbol \eta_i = N \boldsymbol \Upsilon$. %at each iteration after sampling, regardless of whether the regression component is included. 
The $\boldsymbol \phi_i$'s are then obtained from applying the inverse Jupp transformation to the centered $\boldsymbol \eta_i$'s. %The parameter $\rho$ is defined with stringent conditions, which ensures no loss of identifiablity in the realm of the time-transformation functions while allowing for additional flexibility at the population level.

A typical source of non-identifiability in mixed membership models is related to the rescaling invariance of the membership components \citep{Chen2023}. This problem can be addressed by requiring that the mixing probabilities satisfy the \textit{separability condition}, which requires at least one observation to fully belong to each cluster and that can be relaxed to the \textit{sufficiently scattered condition}, a set of geometric conditions. \citet{Chen2023}'s work in the context of topic models provides an overview of the literature addressing these two conditions. \citet{Marco2024} address those concerns for functional mixed membership models, and develop the Membership Rescale algorithm for the two-feature case. Our semi-supervised setting automatically addresses those issues when observation are a priori assigned to each of the clusters. In the case where only  one cluster has observations labelled to it, we apply the following transformation to the $\boldsymbol \pi_i$'s
$$
    \tilde \pi_{i} = \frac{\pi_{i} - \min_j \pi_{j} }{\max_j \pi_{j}- \min_j \pi_{j}}, \quad  \forall i \in \{1, ..., N\},
$$
which ensure that the separability condition is satisfied.

Furthermore, we address additional concerns surrounding the functional features by adopting a semi-supervised approach. That is, we assign a fixed number of the subjects to fully belong to clusters (see \ref{fig:propsup} for an analysis of model sensitivity to the proportion pre-assigned). Our model allows for the scenario where subjects are assigned a priori to either one or both clusters. We explore the performance of both options in simulations and data analysis, and describe a heuristic approach for determining which subjects are labelled for the EEG signals case study. The semi-supervised approach allows to mitigate the label-switching problem. Relabeling algorithms, such as the method of \citet{Stephens2000} that is founded on decision theoretic arguments, are an alternative solution to this problem. However, the semi-supervised approach has the added benefit of clusters being assigned scientific meaning. Additionally, it ensures that the overall shape of the features is identified, as otherwise the model may converge to different combinations of membership components and feature shapes that additively yield comparable fits for each of the observations (see Figure in appendix). We borrow from the Generalized Additive Model literature \citep{Wood2018} and include sum to zero constraints to fix the level of the curves, by setting $\boldsymbol \gamma_1 ^\prime \boldsymbol 1 = 0 $.

\subsection{Posterior Inference}\label{postinf}
\subsubsection{MCMC Implementation}
The full parameter vector of interest is $$
\boldsymbol \theta = (\{c_i\}_{i=1}^N, \{\boldsymbol{\pi}_i\}_{i=1}^N, \{\boldsymbol \phi_i\}_{i=1}^N, \{\boldsymbol \gamma_k\}_{k=1}^K, \rho, \sigma^2_\epsilon, \sigma^2_c, \sigma^2_\phi, \{\lambda_k\}_{k=1}^K)^\prime.
$$
We conduct inference using the Metropolis-within-Gibbs algorithm. The use of conjugate priors allows for closed form posterior conditional distributions for most components of $\boldsymbol \theta$, which are provided in the supplementary material. The mixed membership components $\boldsymbol \pi_i$, the time-transformation spline coefficients $\boldsymbol \phi_i$ and the time-transformation rescaling parameter $\rho$ however require a Metropolis-Hastings step. 

We use a Dirichlet proposal distribution for the $\boldsymbol \pi_i$, centered at the previous value and scaled by a large constant $a$, which we set to be equal to 1000. They are sampled individually, and rescaled at the end of each iteration, if needed. New $\tilde {\boldsymbol{\eta}}_i$ values are proposed from a Gaussian distribution centered at the previous value, with identity covariance scaled by $\tau_i$. The acceptance rate is controlled by updating the value of $\tau_i$ at each iteration as follows 
$$
\tau_i \leftarrow \tau_i \Big(1 + \frac{\text{current acceptance rate - target acceptance rate}}{\sqrt{\text{current iteration number}}}\Big).
$$
In practice, we set the target acceptance rate to 0.35. The $\boldsymbol \phi_i$ also require to be updated individually. The Metropolis-Hastings steps of the sampler as a result represent the largest computational burden. Finally, the Metropolis-Hastings step for the scale of the second feature time-transformation proposes a new value for $\rho$ from a truncated normal distribution with $[0,\infty] $ support, centered at the previous value of $\rho$ and with 0.01 standard deviation.

\subsubsection{Posterior Inference}
We now describe how posterior inference is conducted, specifically how the posterior distributions for parameters (or functions of the parameters) are obtained in our applications.  We first describe how to obtain the posterior distribution for the fit of the curves. Letting $j = 1, ..., J$ index the MCMC samples, the fit for curve $i$ at iteration $j$ is given by
\begin{equation}
    y_i^{(j)}(t) = c_i^{(j)} + \pi_{i}^{(j)}\boldsymbol {\mathcal B}_s(\boldsymbol {\mathcal B}_w(t)^\prime \boldsymbol \phi_i^{(j)})^\prime \boldsymbol \gamma_1^{(j)}+ (1-\pi_{i}^{(j)})\boldsymbol {\mathcal B}_s(\boldsymbol {\mathcal B}_w(t)^\prime \boldsymbol \phi_i^{(j)})^\prime \boldsymbol \gamma_2^{(j)}.
\end{equation}
In our applications,  we evaluate the functional value of the curve at the observed time points $t_{i,1},..., t_{i,n_i}$. At each of the $n_i$ time points, we obtain then obtain a posterior distribution based on $J$ samples, and summaries such as the mean or median can then be computed to obtain a posterior estimate for the fit. 

Within the curve registration framework, recovery of the warping function $h_i ^{-1}(t)$ and  subsequent registering of curves is typically the principal task of interest. For each subject $i$, we obtain $J$ samples of the time-transformation function at time $t$
\begin{equation}
    h_i^{(j)}(t) = \boldsymbol {\mathcal B}_w(t)^\prime \boldsymbol \phi_i^{(j)} .
\end{equation}
The samples can be used to obtain the posterior summary of interest (e.g. mean or median), which we denote by $\hat h_i(t)$. In practice, we evaluate it at the discrete observation times $t_{i,1}, ..., t_{i, n_i}$, such that \begin{equation}
    \hat h_i(t_{i,l}) \equiv \tilde t_{i,l} \implies  \hat h_i^{-1}(\tilde t_{i,l}) = t_{i,l}, \quad l =1, ..., n_i.
\end{equation}
The registered curve for individual $i$ is given by 
\begin{equation}
    Y_i^*(t) = Y_i(t) \circ h_i(t),
\end{equation}
which then translates into the discretely observed data as
\begin{equation}
    Y_i^*(\tilde t_{i,l}) = Y_i( t_{i,l}), \quad l = 1,..., n_i. 
\end{equation}
   
The procedure described disregards the rescaling of the time-transformation effect for the second feature. While it is impossible to align the data fully when $\rho$ is included in the model, one can register with respect to each feature distinctly, using the samples $\rho ^{(j)}(\boldsymbol {\mathcal B}_w(t)^\prime \boldsymbol \phi_i^{(j)}-t) + t$ to obtain the warping function associated with the second shape. We note that if one wishes to apply the method as pre-processing step to obtain a registered sample, it may be preferable to operate in the framework where $\rho =1$. The model is however primarily developed with the motivation of recovering meaningful characteristics of the data. In this context, making the assumption that both features are warped by an identical mechanism is unjustified.

Recovery of the common shape functions $f_k$ is another key component of our analyses. We obtain draws from the marginal posterior distribution of the spline coefficients $\boldsymbol \gamma_k ^{(j)}$. In our analysis, we then typically work with a posterior estimate $\hat {\boldsymbol\gamma}_k$, such as the mean or the median, that can be plugged in to obtain $\hat f_k(t)= \boldsymbol {\mathcal B}_s(t)^\prime\hat {\boldsymbol\gamma}_k$, $t \in \mathcal{T}$. For performance analysis, we evaluate the shape functions at the observed time points, but consider a finer grid for visualization in order to obtain smooth curves. $100(1- \alpha)\%$ credible intervals can be obtained by ordering the sample such that $\tilde{\boldsymbol \gamma}_k ^{(1)}< ...< \tilde{\boldsymbol{\gamma}}_k^{(J)} $, and we have that $\text{CI}_\alpha = [\tilde{\boldsymbol \gamma}_k ^{(\alpha/2)}, \tilde{\boldsymbol \gamma}_k ^{(1- \alpha/2)}]$. Plugging in the quantiles we can then obtain credible intervals for the whole function. The same approach can be adopted to obtain credible intervals for any parameter, or function of parameters. 

As part of our case study, detailed in Section~\ref{casestudysec}, we wish to draw inference on the location of the peak alpha frequency. Exploiting the semi-supervised nature of the analysis, we enforce that the first feature encodes the PAF shape. At each iteration, we obtain a posterior draw $f_1^{(j)}(t) =\boldsymbol {\mathcal B}_s(t)^\prime{\boldsymbol\gamma}_1^{(j)} $. Evaluating the function on a fine grid of values of $t \in \mathcal{T}$, we numerically find the location of the maximum, corresponding to the PAF. This yields a sample of size $J$ from the posterior distribution of the PAF.

\section{Applications} \label{applications}
The simulation study and case study analyses are run in \texttt{R} package and the code is available in the package \texttt{crMM}, available on Github (put link). The model was found to be sensitive to prior choices, specifically with regards to the time-transformation functions. A careful tuning of hyperparameters was conducted for the analysis of the case study to ensure convergence of the chain; the values were found to yield good performance in the simulation context as well. 

\subsection{Simulation Study: Assessing Recovery of the Truth} \label{simulationstudy}

\begin{figure}[h!]
    \centering
        \includegraphics[width=0.48\textwidth]{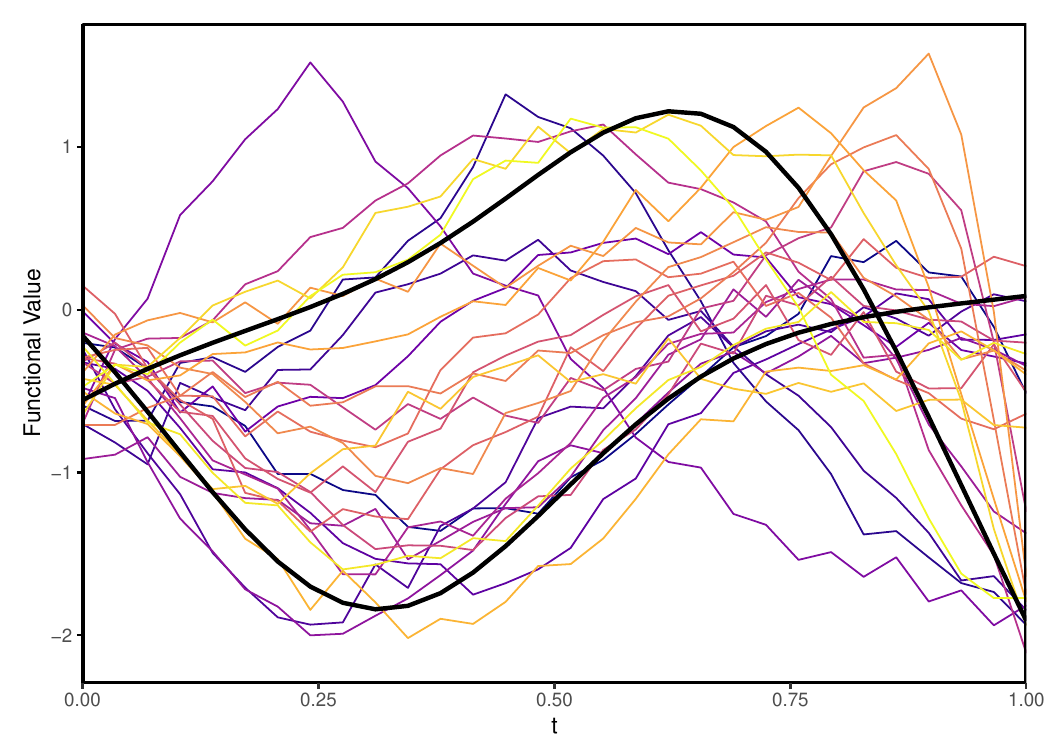} 
    \caption{Simulation study data. Twenty-five simulated at $n= 30$ time points between equally spaced between 0 and 1. The bold black lines represent the shapes associated with the two features.}
    \label{fig:simdata}
\end{figure}

Before proceeding to data analysis of the EEG observations, we run a simulation study to assess the performance of our model. In particular, we focus on the recovery of the population parameters -- namely the shape functions and the rescaling parameter $\rho$.

We consider datasets with $n = 30$ observations per subject at equally spaced time points on $\mathcal T = [0,1]$. We set $p = 15$ to be half the number of observations, and hence have an $R = 19$-dimensional cubic B-spline basis for the shape functions. We set $h= 1$, yielding a $Q =5$-dimensional cubic B-spline basis for the time-transformation functions. In both cases, we assume the interior knots to be equally spaced on the time domain. We set $$
f_1(t) = \Phi\Big(\frac{t- 0.7}{0.2} \Big) - 5 (t- 0.3) ^2 + \text{constant}, \quad f_2(t)= - \Phi\Big (\frac{t- 0.3}{0.2}\Big) + (t-0.7)^2 ,$$
where $\Phi(\cdot)$ denotes the CDF of the standard Normal distribution. We obtain the values of $\boldsymbol \gamma_1$ and $\boldsymbol \gamma_2$ by evaluating the two functions at 101 equally spaced points on $[0,1]$ through simple B-spline regression, with the parameters specified previously. A constant is added to $f_1$ to ensure that the entries of $\boldsymbol\gamma_1$ sum to 0, to help establish identifiability between the simulated data and the sampling model. The simulated curves are then of the form $$
y_{ij} = c_i + \pi_{i,1} \boldsymbol {\mathcal B}_s(\boldsymbol {\mathcal B}_w(t_j)^\prime \boldsymbol \phi_i)^\prime \boldsymbol \gamma_1 + (1 -\pi_{i,1}) \boldsymbol {\mathcal B}_s(\rho(\boldsymbol {\mathcal B}_w(t_j)^\prime \boldsymbol \phi_i - t_j) +t_j )^\prime \boldsymbol \gamma_2 + \epsilon_{i,j}.
$$
We simulate $c_i \stackrel{\text{iid}}{\sim} \mathcal{N}(0; \sigma_c^2 = 0.2^2), \boldsymbol{\pi}_i \stackrel{\text{iid}}{\sim} \text{Dirichlet}(0.5, 0.5)$, $\boldsymbol {\tilde \eta}_i \stackrel{\text{iid}}{\sim} \mathcal{N}_{3}(\boldsymbol {\tilde \Upsilon}, \sigma^2_\phi (=1.5) \mathbf I)$ (and $\boldsymbol{\phi}_i$ obtained appropriately with the inverse Jupp transformation), and $\epsilon_{ij}\stackrel{\text{iid}}{\sim} \mathcal{N}(0; \sigma_\epsilon^2 = 0.085^2) $. We fix $\rho = 0.4$. For proper identification of the parameters, we center $c_i$ and $\boldsymbol{\eta}_i$ as described in section~\ref{identifiability}. The model does not include the regression component for the time-transformation mean structure. Furthermore, we operate in the semi-supervised setting where we assign certain subjects to fully belong to feature 1. We set the 5\% largest $\pi_{i,1}$'s to equal 1, and then rescale to address the separability condition. A simulation of twenty-five curves from the generative model described is illustrated in Figure~\ref{fig:simdata}. 

The central focus of the simulation study is examining the performance of the model as sample size increases. We consider $N = 50, 100, 150$, and in each case simulate 40 datasets on which the model is ran. The hyperparameters for the inverse-gamma priors are chosen to be $a_\epsilon = 0.0001, b_\epsilon = 0.0001$; $a_c = 0.1, b_c = 1$; $a_\lambda = b_\lambda = 0.01$ and $a_\phi= 100, b_\phi = 1$. We set $\boldsymbol \alpha = (0.5, 0.5)$ as the hyperparameter for the Dirichlet distribution. We assign $5\%$ of the observations to fully belong to feature 1. The inference is then based on a 15,000 samples from the posterior distribution, after a burn-in of 45,000 samples.

\begin{figure}[h!]
    \centering
        \includegraphics[width=0.48\textwidth]{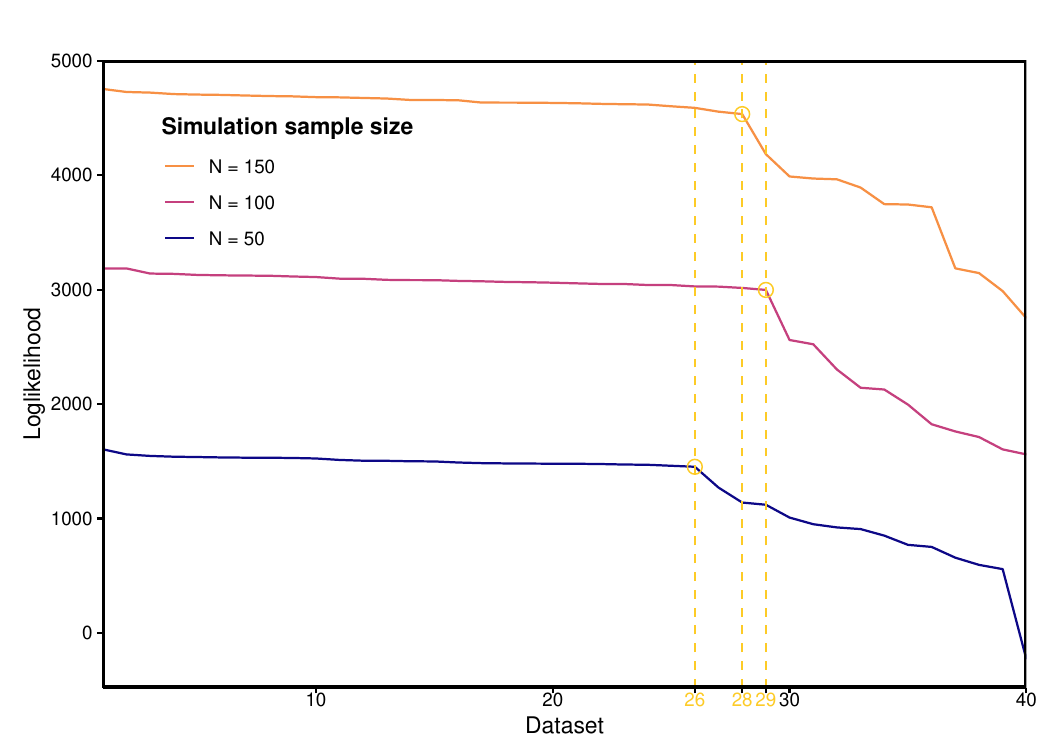} 
    \caption{Median loglikelihood for the 40 datasets of each sample size, ordered by decreasing likelihood. The index of the dataset at which the loglikelihood displays a drop is highlighted.}
    \label{fig:loglik}
\end{figure}

The results from the simulation study hint towards multimodality of the posterior. Figure~\ref{fig:loglik} displays the median loglikelihood of the posterior sample for each of the datasets, represented in order of decreasing values. For all sample sizes, we find the loglikelihood to remain at high values for most of the datasets, apart from certain ones for which it drops. An inspection of those samples reveals convergence towards biased values, in particular of $\rho$, which in turn can lead to inaccurate estimates of quantities such as the shape or the time-transformation functions. The multimodality of the posterior is not an uncommon issue in the high-dimensional Bayesian modelling of functional data. For instance, \cite{Marco2024} consider a tempered transition scheme and tuning of the starting values to address those concerns. For the purposes of the simulation study, we choose to restrict our analyses to the samples that converged to the high likelihood mode. As identified in Figure~\ref{fig:loglik}, this implies working with 26 datasets for $N=50$, 29 datasets for $N= 100$ and 30 datasets for $N=150$.

To measure the accuracy of our estimates, we use the mean squared error (MSE) between our estimate $\hat \rho$ (taken to be the posterior median) and its true value. With regards to the function estimates, we use the relative mean integrated square error (R-MISE), defined as \begin{equation}
    \text{R-MISE} = \frac{\mathbb E[\int (f(t)- \hat f(t)^2)dt]}{\int f(t)^2 dt}, 
\end{equation}
where $\hat f$ denotes the estimated function. We take  $\hat f$ to be the function obtained from the posterior median estimates of the spline coefficients. In practice, we compute \begin{equation}
    \widehat{\text{R-MISE}} = \frac{\sum (f(t_j) - \hat f(t_j))^2}{\sum f(t_j)^2},
\end{equation}
where $t_j$ are 30 equally spaced time points in $[0,1]$.

\begin{figure}[h!]
    \centering
    \begin{tabular}{ccc}
        \includegraphics[width=0.31\textwidth]{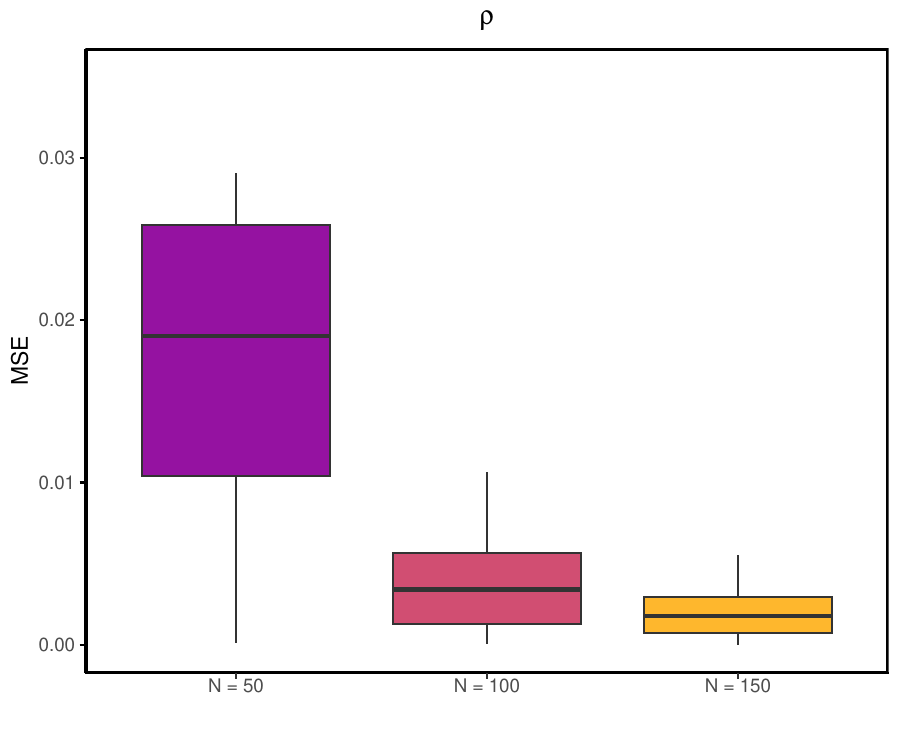} &
         \includegraphics[width=0.31\textwidth]{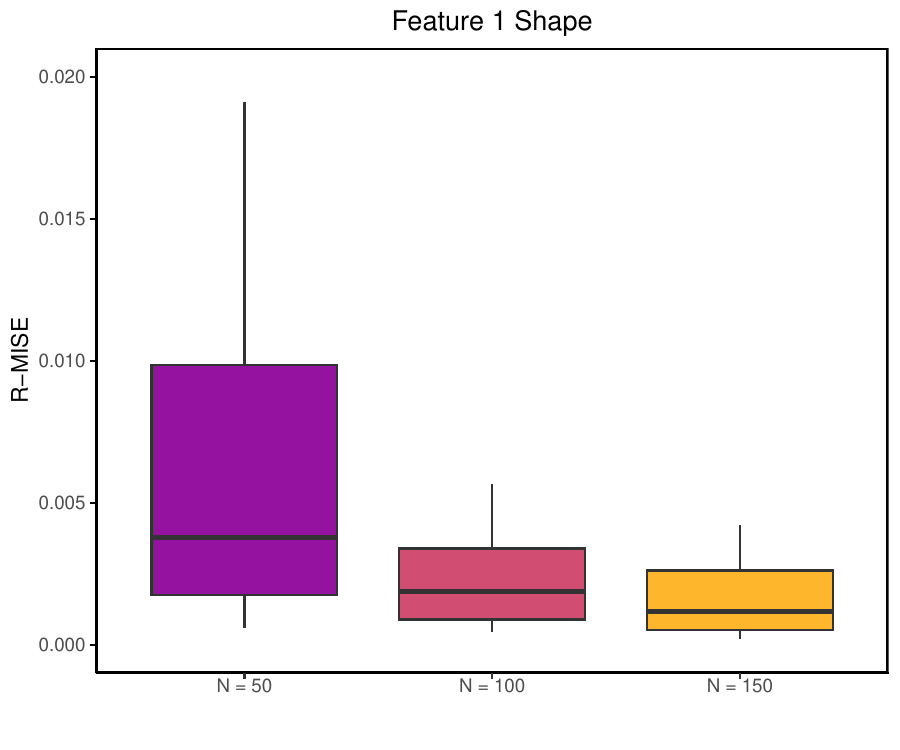} &
         \includegraphics[width=0.31\textwidth]{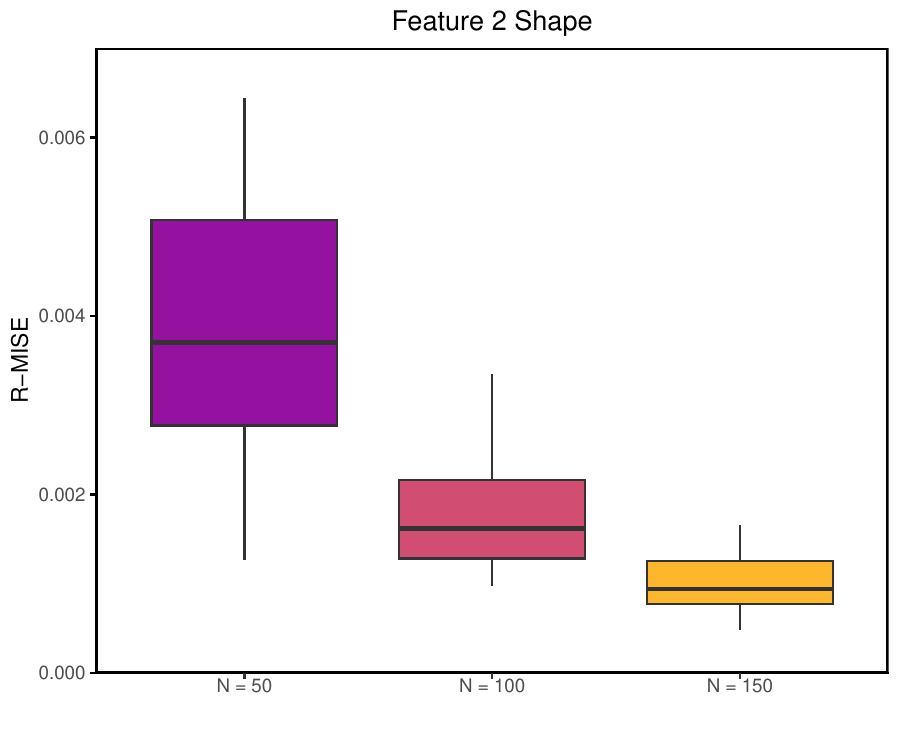}\\
         (a) & (b) & (c)
    \end{tabular}
    \caption[Error of the posterior population parameter estimates as sample size varies.]{Error of the posterior population parameter estimates as sample size varies. The sample sizes considered are $N= 50, 100, 150$ and the boxplots are constructed from running MCMC on 40 simulated datasets. (a) MSE for $\rho$, (b) R-MISE for $\hat f ^*_1(t)$ and (c) R-MISE for $\hat f ^*_2(t)$.}
    \label{fig:samplesize}
\end{figure}

After running the model on 40 datasets, for each of the three considered sample sizes, we find that the posterior estimates of the shape associated with feature 2 display a mismatch in amplitude from the truth. We suggest a post-processing rescaling mechanism applied to the true function and its estimate, defined as follows, 
\begin{align}
    f^*(t) &= \frac{f(t) - \bar f}{\text{SD}(f)}\\
    \hat f^*(t) &= \frac{\hat f(t) - \bar {\hat f}}{\text{SD}(\hat f)}
\end{align}
The mean and standard deviation are obtained across observations, that is $$\bar f = \frac{1}{n}\sum_{j=1}^n f(t_j), \quad \text{SD}(f)= \sqrt{\frac{1}{n-1}\sum_{j=1}^n (f(t_j)- \bar f)^2}, \quad n =30.$$
We provide figures showing the posterior estimates before and after this rescaling in the supplementary material. 

Figure~\ref{fig:samplesize} shows boxplots of the error, allowing to compare accuracy of the model as the sample size increases. The first panel corresponds to $\hat \rho$, the second panel to $\hat f = \hat f^*_1$, and the last to $\hat f = \hat f^*_2$. We find that for all three parameters, the error and the variability around it decrease as the sample size increases. We note that the decrease in R-MISE with sample size for the first shape is not as pronounced. This is coherent with the semi-supervised nature of our model, where the extra constraint allows for reduced flexibility in estimation. Additionally, we find a stark decrease in error between sample sizes $N = 50$ and $N = 100$, but a much lesser difference once it is increased to 150. This is indicative that the model does not require large sample sizes for good performance, and provides reassurance that the case study analysis is reliable (where $N=97$).

% \begin{figure}[h!]
%     \centering
%     \begin{tabular}{ccc}
%         \includegraphics[width=0.31\textwidth]{Simulations_Updated/boxplot_assignments3.pdf} & 
%         \includegraphics[width=0.31\textwidth]{Simulations_Updated/boxplot_assignments1.pdf} & 
%          \includegraphics[width=0.31\textwidth]{Simulations_Updated/boxplot_assignments2.pdf} \\
%         (a) & (b) & (c)
%     \end{tabular}
%     \caption[R-MISE for the posterior shape estimates as the proportion of labeled samples varies]{R-MISE for the posterior shape estimates as the proportion of labeled samples varies. The proportions assigned to feature 1 considered are 1.25\%, 2.5\%, 5\%, 6.25\%, 7.5\%, 8.75\% and 10\%. The boxplots are obtained from running MCMC on 40 simulated datasets of 80 observations. (a)-(c) are R-MISE values for $\hat f = \hat f_1+ \hat f_2, \hat f = \hat f_1$, and $\hat f = \hat f_2$ respectively, computed using raw estimates. (d)-(f) consider the same functions, but applying the rescaling mechanism.}
%     \label{fig:propsup}
% \end{figure}

We leverage the context of our simulation study to also explore the sensitivity of the model to the proportion of subjects that are labelled a priori as fully pertaining to one feature. While the semi-supervised nature is desirable in that it aids with identifiability and is motivated by a desire for scientific interpretability in application, the concern arises that the model's performance becomes overly dependent to these prior assignments. We consider 40 simulated datasets consisting of $N = 80$ functional observations, and vary the proportion of subjects assigned to feature 1 (varying from 1 to 8 labeled observations). We find the errors in recovery of the true parameters to not be impacted by the number pre-assigned, and the model is identifiable. This justifies prior labelling of few observations, which is desirable for maintaining flexibility in the model. 

\subsection{Case Study: Power Spectral Density in the Alpha Band of EEG Signal} \label{casestudysec}
Our motivating case study considers resting-state electroencephalograms (EEGs) measurements obtained from 39 typically developing (TD) children and 58 children with Autism Spectrum Disorder (ASD). They are transformed from the time domain to the frequency domain using the Fast Fourier Transform (FFT), and constrained to the alpha band corresponding to frequencies between 6 and 14 Hz. Measurements were obtained using 25 electrodes that record brain signals from the different regions, however it was found by \citet{Scheffler2019} that the T8 electrode has the highest contribution to the log-odds of ASD diagnosis. In this analysis we will therefore focus on data from this electrode, which corresponds to the right temporal region.

The main feature of interest in the analysis of the alpha band is the peak alpha frequency (PAF), the location of a single prominent peak in the spectral density. It has been found that the PAF shifts to higher frequencies as TD children grow older \citet{Miskovic2015}. For ASD children, the peak may be absent, less pronounced, or display a delayed shift in frequency with age. \citet{Marco2024}, \citet{Marco2024_2} explore this dataset using functional mixed membership models with two features. The first feature is interpreted as $1/f$ noise, while the second feature displays an apparent alpha peak. The loadings to each feature were found to differ significantly between the TD and ASD groups, with higher membership to feature 2 for the TD group. This modelling framework however fails to account for the shifts in peak location, motivating the need for a procedure that includes registration.

\begin{figure}[h!]
    \centering
    \begin{tabular}{c c}
        \multicolumn{2}{c}{\includegraphics[width=0.95\textwidth]{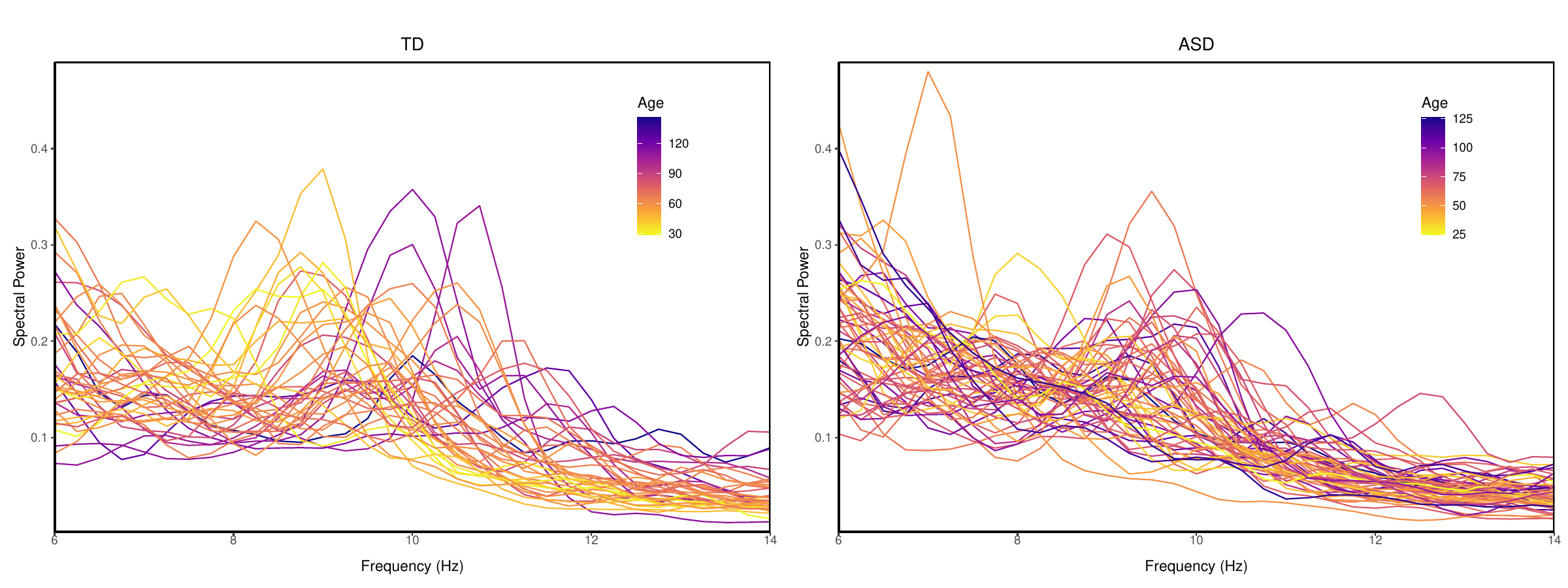}} \\
        \makebox[0.47\textwidth]{(a)} & \makebox[0.47\textwidth]{(b)} \\
        \multicolumn{2}{c}{\includegraphics[width=0.95\textwidth]{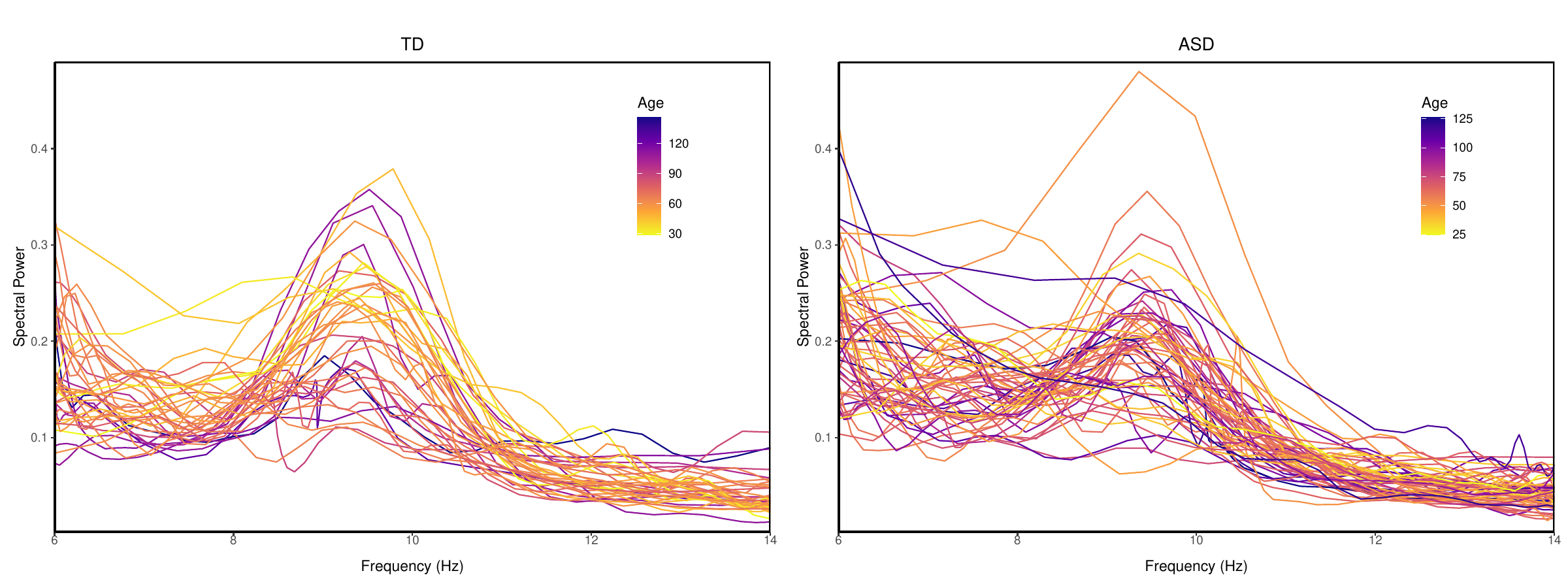}} \\
        \makebox[0.47\textwidth]{(c)} & \makebox[0.47\textwidth]{(d)} \\
    \end{tabular}
    \caption{Spectral density data of the EEG recordings at the T8 electrode, split by clinical designation. Observations are colored by age. (a) and (b) represent the raw data for TD and ASD children respectively, while (c) and (d) display the curves aligned using our model. }
    \label{fig:data}
\end{figure}

Figure~\ref{fig:data} illustrates the dataset. Observations are split by clinical designation, and colored by the child's age in months. The top panels represent the raw, or misaligned data; and the bottom ones display the curves after being registered by our model. Looking at panel (a), the shift in location of the peak towards high frequencies as age increases can be observed amongst TD children, but is not apparent for their ASD counterparts in panel (b). Those remarks, corroborated by the scientific literature, motivate the inclusion of regression in the estimation of the mean structure for the time-transformation functions. 

All results are obtained from a model with assignment of subjects to both features, and with time-transformation regression. The covariates that are included are age, clinical designation, and their interaction. We set the number of interior knots to $p=15$ for the shape coefficients, and $h = 1$ for the time-transformation functions. The hyperparameters are $a_\epsilon = b_\epsilon = 0.0001$; $a_c = 0.1,  b_c = 1$; $a_\lambda = b_\lambda = 0.01$, $a_\phi= 100, b_\phi = 1$ and $\boldsymbol \alpha = (0.5, 0.5)$, as in the simulation study. We set $g = 97$, the sample size, for the covariance in the g-prior. Posterior inference is conducted on 15,000 samples, after removing 85,000 MCMC iterations for burn-in. 

We apply our model to the data, re-scaled with the $\log (y+1)$ transformation, in a semi-supervised fashion.  We impose that 5\% of the subjects are labeled; in the case of our data this implies imposing that two individuals are set to have $\pi_{1} = 1$ and two to have $\pi_{2}=1$. The set of individuals chosen to belong fully to the second feature are those who display the most prominent peak. We select those by computing the difference in spectral power between the most significant peak found in the data (if applicable) and its base. The ones with the largest value are assigned to fully belong to the second cluster. On the other hand, the set of individuals chosen to belong to the first feature are those that display pure $1/f$ noise, selecting the observations that do not display a peak. To do so, we regress the functional spectral power measurements on frequency, restricted to the 7-12 Hz domain. The observations with the lowest sum of squared errors are selected, as we expect the second feature to reflect a decreasing trend, without the presence of a peak that would lead to poor linear fit. Pre-labeling subjects to only one of the features was explored, but was found to cause issues with convergence of the shape functions.

Issues were found to arise with the estimation of the parameter $\rho$. The posterior appeared to display two modes: one for a value close to 0, and one for a value close to 1. The sampler eventually converges to the latter, in which case the shape associated with the $1/f$ noise would also display a small peak that could get warped. This leads to issues with identifiability of the time-transformation function and the mixed membership component, in particular for subjects that may display two small peaks. As a result, $\rho$ was set to be equal to 0 in the context of the case study, rather than being estimated.

\begin{figure}[h!]
    \centering
    \begin{tabular}{cc}
        \includegraphics[width=0.48\textwidth]{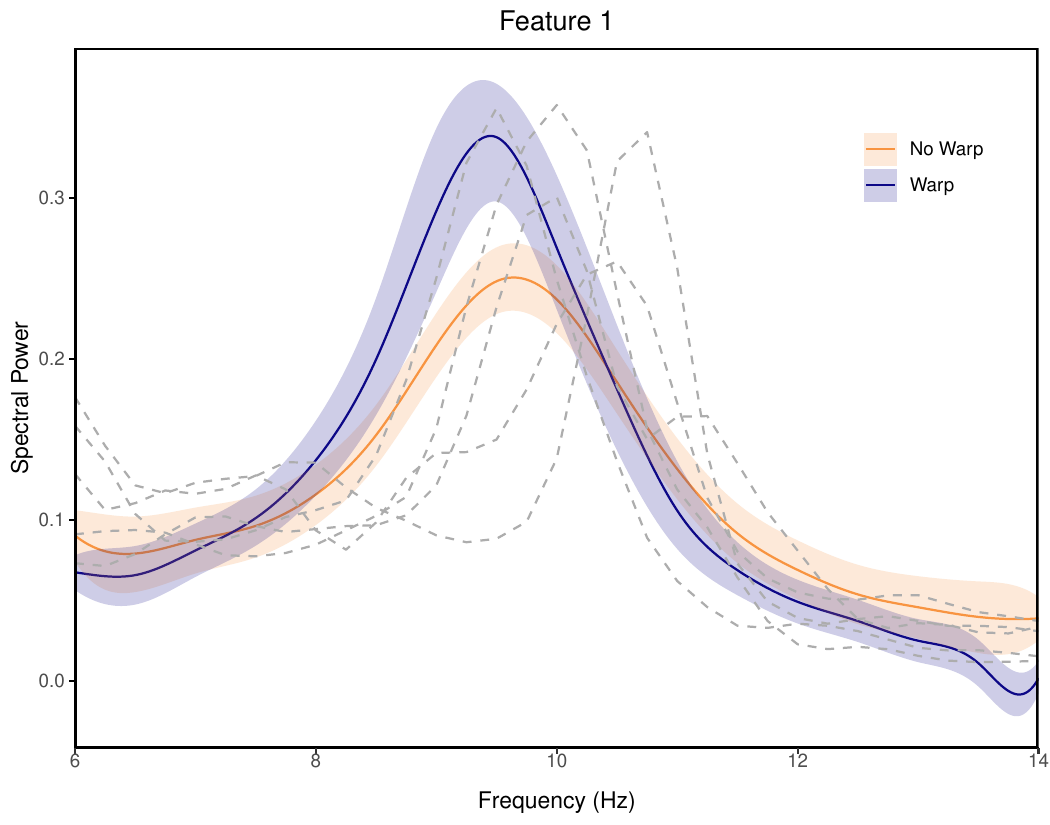} & \includegraphics[width=0.48\textwidth]{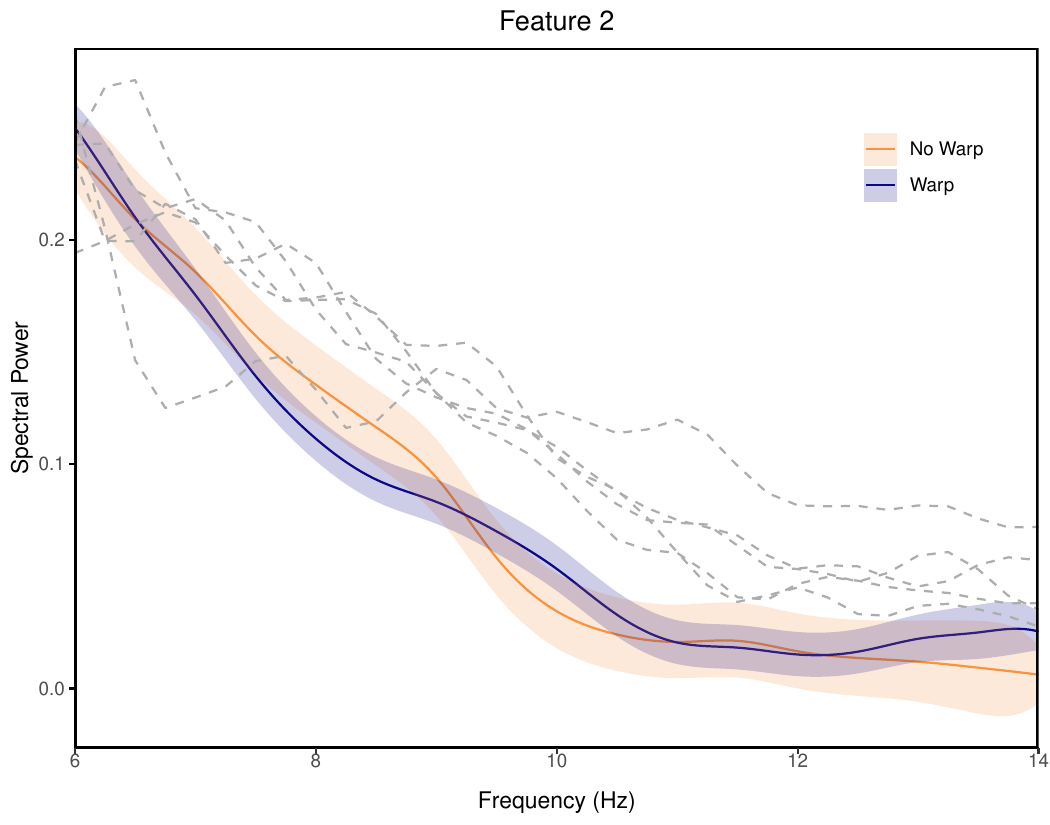} \\
        (a) & (b) 
    \end{tabular}
    \caption{Comparison of the individual feature shapes recovered between a warping and no warping model. The solid line represents the posterior median, while the shaded area is the 95\% credible interval. In panel (a), the estimates for the first feature are superimposed on the five observations with the most prominent peak in (a). In panel (b), five observations with no apparent peak are displayed.}
    \label{fig:shapes}
\end{figure}

The shapes recovered for the model associated with each feature are displayed in Figure~\ref{fig:shapes}. The posterior median for each function are represented jointly with the 95 \% credible interval, estimated for the registration model presented in this manuscript, as well as a comparison to model that does not assume warping (i.e. $\phi_i (t) = t$ for any $i$ and $t$). Looking at panel (a), corresponding to feature 1, we find that accounting for warping allows the model to identify a peak of larger amplitude than with the model that doesn't. In particular, it can be seen to accurately reflect the peak height observed in some of the most extreme observations. On the other hand,  the model without warping recovers a dampened and wider peak that isn't consistent with the observed data, a common issue in FDA. This result highlights the necessity of accounting for phase variability in our case study analysis. Regarding the $1/f$ noise, we find the recovered shape functions to not differ much between the warping and no warping models, which can be expected as it is assumed that $\rho = 0$. We find however that the uncertainty surrounding the estimate is greater for the model that not account for phase variation.

\begin{figure}[h!]
    \centering
         \begin{tabular}{c c}
        \multicolumn{2}{c}{\includegraphics[width=0.95\textwidth]{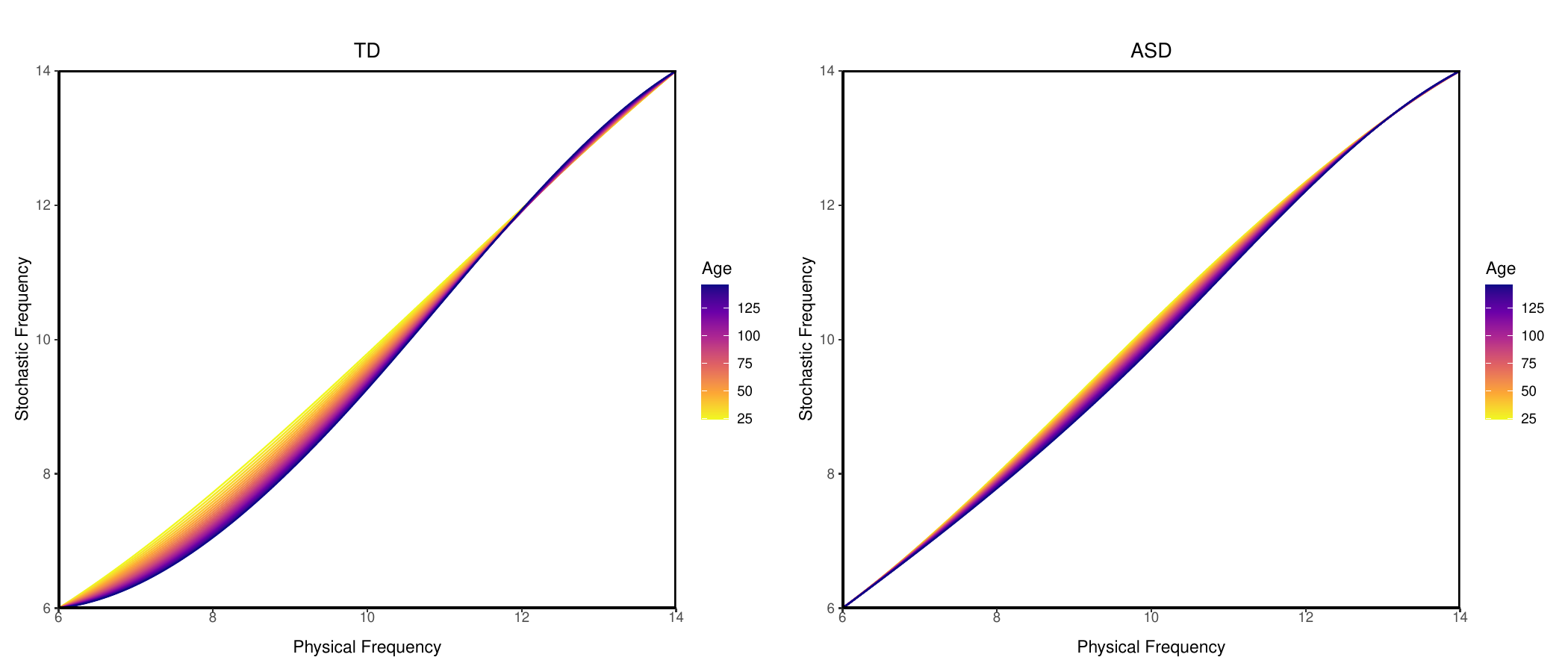}} \\
        \makebox[0.47\textwidth]{(a)} & \makebox[0.47\textwidth]{(b)} \\
        \multicolumn{2}{c}{\includegraphics[width=0.95\textwidth]{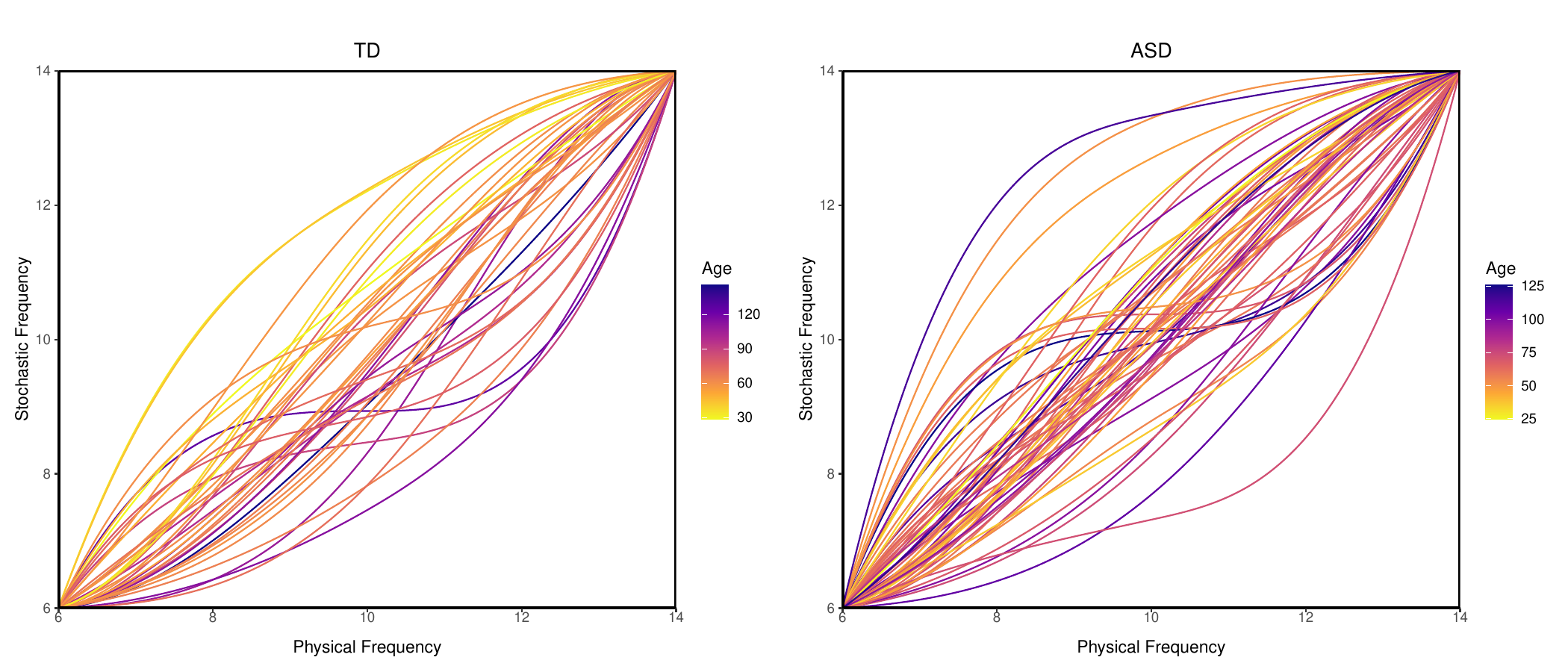}} \\
        \makebox[0.47\textwidth]{(c)} & \makebox[0.47\textwidth]{(d)} \\
    \end{tabular}
    \caption{Posterior time-transformation function analysis. Time-transformation functions are regressed on age in months, clinical designation, and their interaction (a)-(b) Posterior median for the time-transformation mean, split by clinical diagnosis. The color scales such that it represents the posterior regression mean at different ages. (c)-(d) Posterior median for the time-transformation functions, split by clinical diagnosis.}
    \label{fig:regression}
\end{figure}

In the remainder of this section, we explore how the proposed method allows to gain insights into the study of ASD using spectral-transformed EEG data. Figure~\ref{fig:regression} illustrates the posterior median estimates of the time-transformation functions, which relate to the age-related trajectory of the PAF. The top row displays the regression mean, showing how it varies with age, and is split by clinical designation. The regression results for TD children show that, as age increases, the location of the peak gets delayed towards higher frequencies, unless the peak appears beyond 12Hz after which the effect of age is not significant. On the other hand, the mean stays very close to the identity regardless of age within the ASD group. We find those results consistent with previous work studying PAF as a neuromarker for ASD: \citet{Edgar2015} find no significant relationship between age and the location of the alpha peak, whereas it is well documented the alpha peak moves towards higher frequencies with age in TD children. Panels (c) and (d) represent the median posterior time-transformation functions, colored by age and split by clinical designation as above. For the TD group, subjects of younger ages appear to display time-transformation functions above the identity and older ones below, corresponding to the regression findings. No such structure appears for the ASD children. Moreover, we notice much larger variability in the time-transformation functions for the latter group, indicating that the peak location is scattered across the entirety of the alpha frequency domain. The PAF location doesn't deviate as much amongst TD children. 

\begin{figure}[h!]
    \centering
    \begin{tabular}{cc}
        \includegraphics[width=0.47\textwidth]{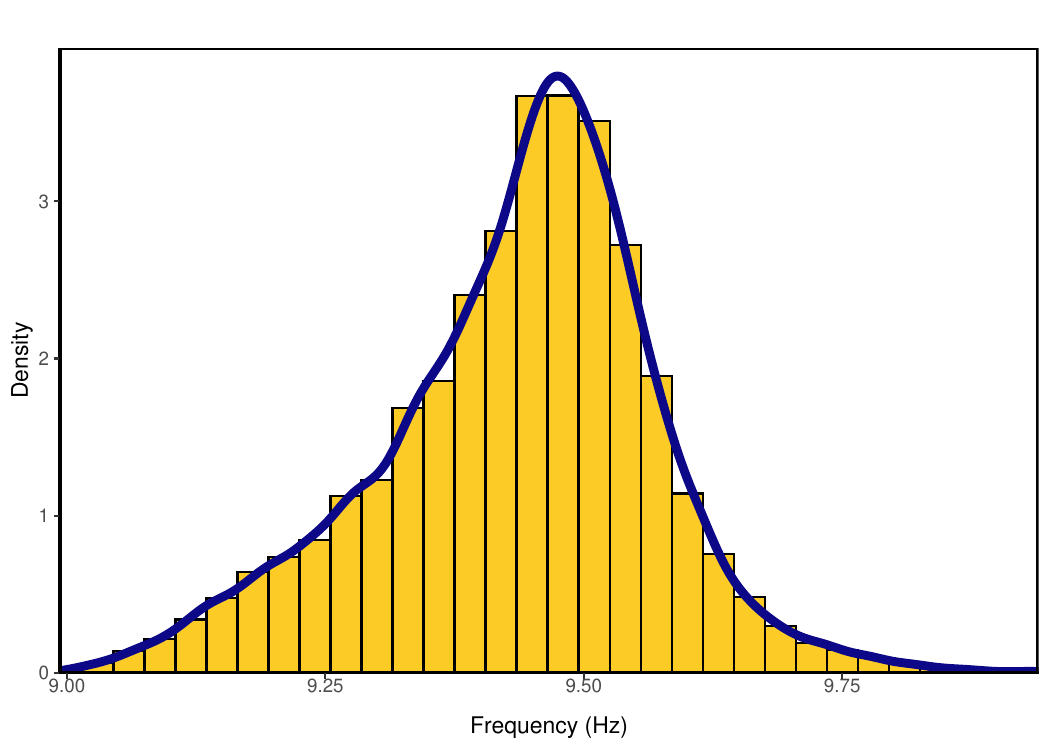} & \includegraphics[width=0.47\textwidth]{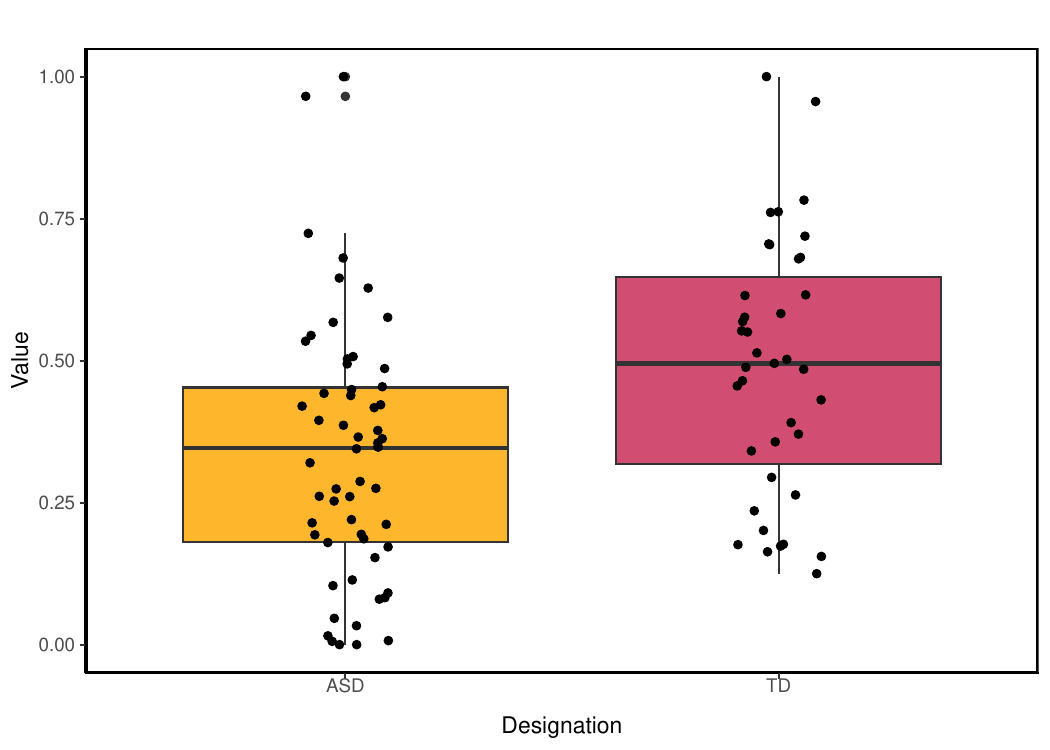} \\
        (a) & (b) 
    \end{tabular}
    \caption{(a) Posterior distribution of the peak alpha frequency location. (b) Posterior mean of individual membership to the first feature $\pi_{i,1}$, stratified by clinical designation.}
    \label{fig:pafmemb}
\end{figure}

While the time-transformation functions inform individual departures from a common shape, the Bayesian approach allows to also conveniently identify the centered PAF location. Its posterior distribution is displayed on the left of Figure~\ref{fig:pafmemb}. We find the distribution to have highest mass in the 9.4-9.5 Hz range, and assigns non-zero probability to values between 9.0 and 9.95 Hz. \citet{Miskovic2015} identified that the PAF location in TD children ranges from 8.89 Hz at age 7 to 9.79 Hz at 11 years of age. Our data is collected on children between 2 and  12 years old, hence our results are coherent with the neuroimaging literature.

Finally, we compare the level of membership to each feature across groups. The right side of Figure~\ref{fig:pafmemb} shows a boxplot of the posterior medians of the individual membership level to feature 1. The TD group is found to load more heavily on it, with a median value of 0.50 across individuals, versus a median of 0.35 for the ASD group.  This implies that the TD group will on average have the curve more strongly governed by the peak, as opposed to the the ASD group for whose observations the shape is dominated by the $1/f$ noise. Our study hence provides evidence that the PAF is less prominent in ASD children-- consistent with the results of \citet{Dickinson2018}, and helps quantify the difference across groups. 

\section{Discussion} \label{discussion}
In this manuscript, we have introduced a Bayesian curve registration method for mixed membership models. For identifiability purposes, we have proposed a semi-supervised framework with partial labelling of observations, that also aids with scientific interpretability of results. Furthermore, estimation of the time-transformation functions in a tangent space via the Jupp transformation provided a convenient framework for incorporating covariate information. Bayesian modelling allows to easily quantify uncertainty of estimates and to obtain posterior summaries of a wide range of quantities of interest. 

Our simulation study demonstrated that the accuracy of estimates for population parameters improves for larger sample size. It also showed that the model is not overly sensitive to the proportion of labelled samples, which justifies fixing a very small proportion of subjects to specific clusters, keeping our model flexible. We then applied the method to real data in the context of the EEG case study. Accounting for phase variation allowed to recover meaningful underlying shape functions for the clusters that represent features of the data. Furthermore, we found that TD subjects were more likely to display an apparent peak alpha frequency, agreeing with the neuroscience literature \citep{Dickinson2018}, and our method quantifies the effect of age and clinical designation on the timing of features. 

Our work is related to the combined factor analysis and curve registration model introduced by \citet{Earls2017}. They also allow for functional variation across two shapes, but their approach differs from the mixed membership point of view. Indeed, in that model data is assumed to arise from a main functional direction, while a weakly weighted second function is include to account for individual departures from the primary shape. Our approach allows (and in fact, assumes) that certain subjects may belong fully to any of the subpopulations. The mixed membership vector indicates the individual degree of belonging to each cluster, whereas the factor analysis construction does not have such an interpretation. 

We now discuss certain limitations of our model. Motivated by the case study, this manuscript focuses on the cases where the number of features is known and set to be equal to be two. The method would need to be adjusted to be applicable to more general scenarios where interest may be in recovering $K$ subpopulations.  This number may be unknown, and one would then need to learn its value. This could either be done heuristically, for instance using the "elbow-method" for a chosen information criterion \citep{Marco2024}, or probabilistically by incorporating a prior for $K$ into the model. Another concern is computational efficiency-- the Metropolis-within-Gibbs sampler is costly, in particular since individual level parameters for the time-transformation functions and mixed membership vector are updated one at a time. This could become an issue for larger datasets. 

The model could be extended in different directions. First, we may consider a more sophisticated prior on the mixed membership components, to help tackle nonidentifiability and the \textit{rescaling problem}. For instance, we could instead assume the Determinantal Point Process (DPP) repulsive prior \citep{Xu2016}.  A repulsive covariance structure for the spline coefficients across features could also be considered instead of the current choice of a block diagonal matrix, ensuring the model learns shapes that are different, or far away, from each other. These modifications could eliminate the need for the semi-supervised setting, adding flexibility to the model. We could also consider a multivariate extension to the model. For instance, in the case of the EEG dataset, this would allow the borrowing of information across electrodes. 

\begin{funding}
This work was supported in part by NIH Grant NIMH 5R01MH122428-05R1.
\end{funding}

%%%%%%%%%%%%%%%%%%%%%%%%%%%%%%%%%%%%%%%%%%%%%%
%% Supplementary Material, including data   %%
%% sets and code, should be provided in     %%
%% {supplement} environment with title      %%
%% and short description. It cannot be      %%
%% available exclusively as external link.  %%
%% All Supplementary Material must be       %%
%% available to the reader on Project       %%
%% Euclid with the published article.       %%
%%%%%%%%%%%%%%%%%%%%%%%%%%%%%%%%%%%%%%%%%%%%%%
\begin{supplement}
Supplementary materials is available from the corresponding Author upon request.
%\stitle{Title of Supplement A}
%\sdescription{Short description of Supplement A.}
%\end{supplement}
%\begin{supplement}
%\stitle{Title of Supplement B}
%\sdescription{Short description of Supplement B.}
\end{supplement}

\bibliographystyle{ba}
\bibliography{export}

\end{document}